\documentclass[twocolumn,showpacs,amsmath,amssymb,prl,superscriptaddress]{revtex4}

\usepackage{graphicx}
\usepackage{dcolumn}
\usepackage{bm}

\newcommand{\bs}{\boldsymbol}

\begin{document}

\preprint{APS/123-QED}

\title{An exact relation between Eulerian and Lagrangian velocity increment statistics}

\author{O. Kamps}
\affiliation{Theoretische Physik, Universit\"at M\"unster, Germany}
\author{R. Friedrich}
\affiliation{Theoretische Physik, Universit\"at M\"unster, Germany}
\author{R. Grauer}
\affiliation{Theoretische Physik I, Ruhr-Universit\"at Bochum, Germany}

\date{\today}

\begin{abstract}
We present a formal connection between Lagrangian and Eulerian
velocity increment distributions which is applicable to a wide range
of turbulent systems ranging from turbulence in incompressible fluids to
magnetohydrodynamic turbulence. For the case of the inverse cascade
regime of two-dimensional turbulence we numerically estimate the
transition probabilities involved in this connection. In this
context we are able to directly identify the processes leading to strongly
non-Gaussian statistics for the Lagrangian velocity increments.

\end{abstract}




\maketitle

\begin{figure}[t]
\includegraphics[width=0.4\textwidth]{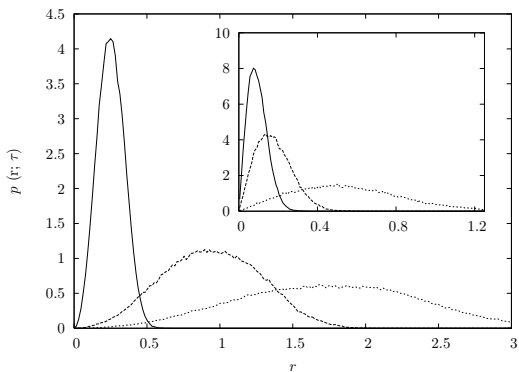}
\caption{The figure shows $p_a(r;\tau)$ for  $\tau =3.5\tau_{\eta}$,
$14\tau_{\eta}$, $28\tau_{\eta}$ for three-dimensional turbulence.
In  the inset $p_a$ is depicted for the two-dimensional case with
$\tau=0.22 T_I,0.44 T_I,0.9 T_I$} \label{fig:diff3D}
\end{figure}

\textit{Introduction} The relation between Eulerian and Lagrangian
statistical quantities is a fundamental question in turbulence
research. It is of crucial interest for the understanding and
modeling of transport and mixing processes in a broad range of research fields
spanning from cloud formation in atmospheric physics over the dispersion of microorganisms in oceans to 
research on combustion processes and the understanding of heat transport in fusion plasmas.
The recent possibility to assess the statistics of
Lagrangian velocity increments by experimental means 
\cite{laporta01nat,mordant01prl} has stimulated
investigations of relations between the Eulerian and the Lagrangian
two-point velocity statistics. Especially, the emergence of intermittency (i.e. the anomalous scaling of the moments of the velocity increment distributions \cite{frisch1995})
in both descriptions and its interrelationship is of great importance for
our understanding of the spatio-temporal patterns underlying turbulence.
A first attempt to relate Eulerian
and Lagrangian statistics has been undertaken by Corrsin
\cite{corrsin1959}, who investigated Eulerian and Lagrangian velocity
correlation functions. Recently, the Corrsin approximation has been
reconsidered by \cite{ott2000}, where it has become evident that it is a
too crude approximation and cannot deal with the question of the
connection between Eulerian and Lagrangian intermittency. 
Further approaches to characterize
Lagrangian velocity increment statistics are based on multifractal models \cite{borgas93,biferale04prl} which also have been extended to the dissipation range \cite{chevillard2003}. In \cite{biferale04prl} a direct translation of the Eulerian multifractal statistics to the Lagrangian picture is presented. In this approach a non-intermittent
Eulerian velocity field cannot lead to Lagrangian intermittency.
This statement is in contrast with the experimental results of
Rivera \cite{rivera2007}, as well as numerical calculations
performed for 2d turbulence \cite{kamps2007}. Motivated by this fact
we have derived an exact relation between the Eulerian and
the Lagrangian velocity increment distributions, which allows to
study the emergence of Lagrangian intermittency from a statistical
point of view.
\\[1ex]
\textit{Connecting the increment PDFs}
The quantities of interest are the Eulerian velocity increments
\begin{equation}
u_e = v(\bs{y}+\bs{x},t) - v(\bs{y},t),
\label{eq:ue}
\end{equation}
where the velocity difference is measured at the time $t$ between
two points that are separated by the distance $\bs{x}$, and the
Lagrangian velocity increment
\begin{equation}
u_l = v(\bs{y}+\tilde{\bs{x}}(\bs{y},\tau,t),t)- v(\bs{y},t-\tau).
\label{eq:ul}
\end{equation}
In the latter case the velocity difference is measured between two
points connected by the distance $\tilde{\bs{x}}(\bs{y},\tau,t)$
traveled by a tracer particle during the time interval $\tau$. In
both cases $v$ is defined as the projection $\bs{v}\cdot
\hat{\bs{e}}_i $ of the velocity vector on one of the axes
($i=x,y,z$ in 3d and $i=x,y$ in 2d ) of the coordinate system 
(see e.g. \cite{mordant04njp,voth2002}). In the case of an isotropic
flow the results do not depend on the chosen axis, however 
we do not have to make this assumption yet. Additionally, we define the
velocity increment
\begin{equation}
u_{el} = v(\bs{y}+\tilde{\bs{x}}(\bs{y},\tau,t),t)- v(\bs{y},t) ,
\label{eq:uel}
\end{equation}
which is a mixed Eulerian-Lagrangian quantity because the points
are separated by $\tilde{\bs{x}}$ but the velocities are measured at
the same time. The properties of this quantity have been
investigated in \cite{friedrich2007}. Finally, we introduce
\begin{equation}
u_{p} = v(\bs{y},t)- v(\bs{y},t-\tau),
\label{eq:up}
\end{equation}
measuring the velocity difference over the time $\tau$ at the starting point
of the tracer. Following \cite{lundgren67pof,pope2000} we define 
the so called fine-grained PDF for $u_e$ as
\begin{equation}
\hat{f}_e(v_e;\bs{x},\bs{y},t) = \delta (u_e-v_e).
\label{eq:fineGrainedUe}
\end{equation}
where $u_e$ is the random variable and $v_e$ is the independent sample-space variable.
The fine-grained PDF describes the elementary event of finding the value 
$v_e$ given the measured  $u_e = v(\bs{y}+\bs{x},t) - v(\bs{y},t)$. The relation to the PDF is determined by
\begin{equation}
f_e(v_e;\bs{x},\bs{y},t)  = \langle \hat{f}_e(v_e;\bs{x},\bs{y},t) \rangle ,
\end{equation}
where the brackets denote ensemble averaging. The quantity $f_e $ is
a function with respect to the variables $\bs{x},\bs{y},t$ 
and a PDF with respect to the variable $v_e$. In analogy to
(\ref{eq:fineGrainedUe}) we can define 
fine-grained PDFs for all other quantities. Now we want to derive an exact relation
between the fine-grained PDFs $\hat{f}_e$ and $\hat{f}_l$. This task
can be split up into two steps. First we have to replace the distance $\bs{x}$ by the trajectory
$\tilde{\bs{x}}(\bs{y},\tau,t)$ of a tracer in order to translate from $u_e$ to
$u_{el}$. This is done by 
\begin{align}
\hat{f}_{el} (v_{e};& \bs{y},\tau,t) \nonumber \\ & = \int d\bs{x}
\ \delta(\tilde{\bs{x}}(\bs{y},\tau,t)-\bs{x}) \hat{f}_e(v_e;\bs{x},\bs{y},t).
\label{fineGrainedUel}
\end{align}
We see that during this operation the sample-space variable is not affected. The subscript in$\hat{f}_{el}$ denotes the fact that we now have $v_e=u_{el}$ instead of $v_e=u_e$.  
In the second step we have to connect $\hat{f}_l$ and $\hat{f}_{el}$. 
From the definitions of the increments (\ref{eq:ul}) - (\ref{eq:up}) 
we see that $u_l = u_{el} + u_{p}$ and therefore the fine-grained PDF 
for $u_l$ is given by the fine-grained distribution of the sum 
of $u_{el}$ and $u_p$. To that end we have to multiply $\hat{f}_{el} (v_{e}; \bs{y},\tau,t)$ with $\hat{f}_{el} (v_{e}; \bs{y},\tau,t)=\delta(u_p - v_p)$ to get the fine-grained joint probability of finding $u_{el}$ and $u_p$ at the same time. Subsequent application of $\int d v_{e} \int d v_p \ \delta (v_l - (v_{e} + v_p))$ leads to
\begin{align}
\hat{f}_{l} (v_{l}&; \bs{y},\tau,t) \nonumber \\ & = \int dv_{e} \ \hat{f}_{p} (v_l - v_{e};\bs{y},\tau,t)  \hat{f}_{el} (v_{e}; \bs{y},\tau,t).
\label{eq:fineGrainedUl}
\end{align}
To derive the corresponding PDFs we have to perform
the ensemble average. In case of equation (\ref{eq:fineGrainedUl}) 
we obtain
\begin{align}
f_l (v_{l}& ; \bs{y}, \tau,t) \nonumber \\ & = \left\langle \int dv_{e} \ \hat{f}_{p} (v_l - v_e;\bs{y},\tau,t)  \hat{f}_{el}(v_e;\bs{y},\tau,t) \right\rangle \nonumber \\ & =  \int dv_e  \ f_{p} (v_l - v_e| v_e;\bs{y},\tau,t)   f_{el}(v_e;\bs{y},\tau,t).
\label{eq:pdfUl}
\end{align}
In the last line we used the general relation 
$p(a,b)=p(a|b)p(b)$ valid for two random variables in order to extract $f_{el}$
from the average. We can treat (\ref{fineGrainedUel})
in a similar manner. This leads us to
\begin{align}
f_{el} (v_e&; \bs{y},\tau,t) \nonumber \\ & = \left\langle \int d\bs{x}
\ \delta(\tilde{\bs{x}}(\bs{y},\tau,t)-\bs{x}) \hat{f}_e(v_e;\bs{x},\bs{y},t) \right\rangle \nonumber \\ &  =
\int d\bs{x} \ \langle 
\delta(\tilde{\bs{x}}(\bs{y},\tau,t)-\bs{x})|v_e \rangle \hat{f}_e(v_e;\bs{x},\bs{y},t)
\label{eq:pdfUel}.
\end{align}
Inserting equation (\ref{eq:pdfUl}) into (\ref{eq:pdfUel}) shows
that the Eulerian and  the Lagrangian velocity increment PDFs are
connected via the transition probabilities $p_a = \langle
\delta(\tilde{\bs{x}}(\bs{y},\tau,t)-\bs{x})|v_e \rangle$ and $p_b =
f_{p} (v_l - v_e| v_e;\bs{y},\tau,t) = f_p
(v_p|v_e;\bs{y},\tau,t)$. Before we connect both
equations we introduce some simplifications. In most
experiments and numerical simulations dealing with Lagrangian
statistics the flow is assumed to be stationary and homogeneous. In
this case we may average with respect to $\bs{y}$ and $t$ and hence the 
dependence on this parameters in (\ref{eq:pdfUel}) and
(\ref{eq:pdfUl}) drops. Under the assumption of isotropy $f_{el}$ depends
only on $r=|\bs{x}|$. Therefore, we can introduce spherical coordinates
in (\ref{eq:pdfUel}) and integrate with respect to the angles. As
mentioned before, in the case of isotropy the statistical quantities
do not depend on the chosen axis. Finally we arrive at
\begin{align}
f_l & ( v_l;\tau) \nonumber \\ & = \int dv_e p_b (v_l-v_e| v_e;\tau) \underbrace{\int_{0}^{\infty} dr \ p_a (r|
v_e;\tau) \ f_e (v_e;r)}_{f_{el}(v_e;\tau)}.
\label{eq:EulerLagrangeIncrement}
\end{align}
\begin{figure}
\includegraphics[width=.4\textwidth]{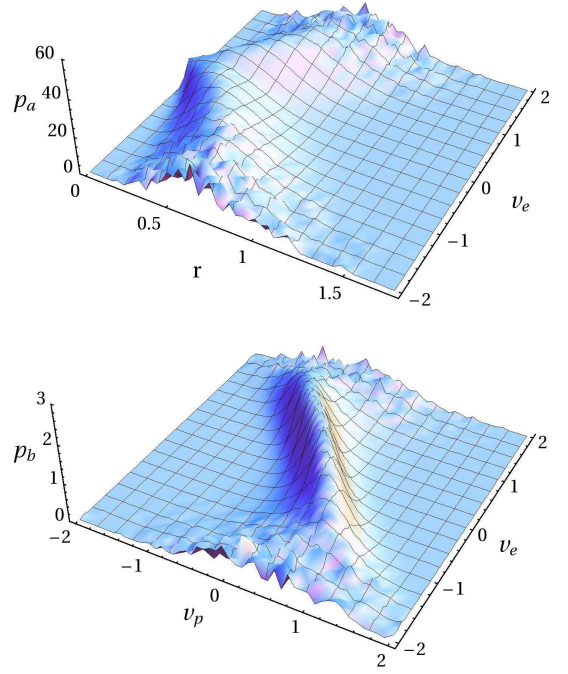}
\caption{The plot shows $p_a(r|v_e;\tau)$ (top) and
$p_a(v_p|v_e;\tau)$ (bottom) for $\tau=0.09T_I$. }
\label{fig:transition}
\end{figure}
For convenience, we included the integrated functional determinant ($2\pi r$ in
two and $4\pi r^2$ in three dimensions ) in $p_a$. In this case $p_a
(r| v_e;\tau)$ is a measure for the turbulent transport and gives
the probability of finding a tracer traveling the absolute distance
$r$ within the time interval $\tau$. Multiplication by $f_e (v_e;r)$ and  subsequent
integration over the whole $r$-range mixes the Eulerian statistics
from different length scales $r$ weighted by $p_a$ to form the PDF
$f_{el}(v_e;\tau)$ for a fixed time-delay $\tau$. The occurence of the condition in $p_a$ shows that this weighting depends on $u_e$. 
changes by $u_p$.
\begin{figure*}
\includegraphics[width=1.0\textwidth]{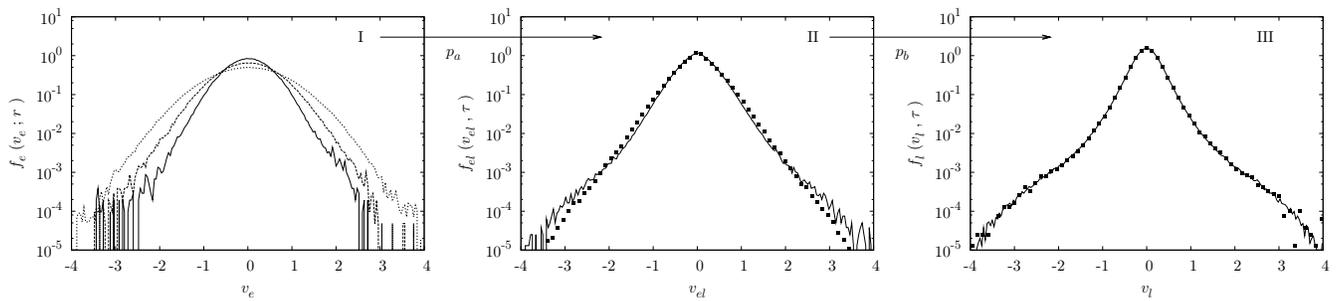}
\caption{The figure shows the impact of the transition probabilities
on the Eulerian PDF. The left part of the figure shows
 $f_e(v_e ; r)$ for $r=0.06,0.12,0.3$. These PDFs are transformed 
into $f(v_e;\tau)$ (middle) by the transition PDF presented in the 
upper half of Fig. \ref{fig:transition}. Subsequently $f(v_e;\tau)$ 
is converted into $f_l(v_l;\tau)$ (right picture) by the second 
transition PDF from Fig. \ref{fig:transition}. 
In both cases $\tau = 0.09 T_I$. In plot II and III 
the points denote the reconstructed PDFs based on
(\ref{eq:pdfUel}) and (\ref{eq:pdfUl}) and the lines denote the PDFs directly computed from the Lagrangian data. 
}
\label{fig:pdfEulerLagrange}
\end{figure*}
The transition probability $p_b$ incorporates the fact that during the
motion of the tracer particle the velocity at the starting point
Multiplying $f_{el}(v_e;\tau)$ with $p_b$ and integrating over
$v_e$ sorts all events where $ u_{el} + u_p = u_l $ into the
corresponding bin of the histogram for $u_l$. 
We want to stress that equations (\ref{eq:pdfUel}) and (\ref{eq:pdfUl}) and except from
symmetry considerations also equation
(\ref{eq:EulerLagrangeIncrement}) are of a purely statistical nature
and, as a consequence, hold for quite different turbulent fields. They are valid
for two-dimensional as well as three-dimensional incompressible turbulence
but can also be applied to magnetohydrodynamic turbulence. 
The differences in the details of these turbulent systems, which are connected to the presence
of different types of coherent structures like localized vortices in 
the case of incompressible fluid turbulence or sheet-like structures in
magnetohydrodynamic turbulence, are therefore closely related to the 
functional form of $p_a$ and $p_b$.
 \\[1ex]
 \textit{Two-dimensional turbulence} In
this section we want to estimate numerically the two transition PDFs
in (\ref{eq:EulerLagrangeIncrement})  for the case of the inverse
energy cascade of two-dimensional turbulence.
The data are taken from a pseudospectral simulation of the inverse 
energy cascade in a
periodic box with box-length $2\pi$ \cite{kamps2007}. Recapitulating the
derivation of
(\ref{eq:EulerLagrangeIncrement}) we see that we need the velocity
at the start and the end point of a tracer trajectory at the
same time to estimate the transition PDFs. Therefore we have to
record the velocity at the starting points of the tracers
additionally to their current position and their current velocity.
In Fig. \ref{fig:transition} the transition probabilities 
$p_a$ and $p_b$ are depicted for $\tau
= 0.09 T_I$, where $T_I$ denotes the Lagrangian integral time scale \cite{yeoung2002}. We have
chosen this rather small time lag as an example because in this case
the deviation of the Lagrangian increment PDF from a Gaussian is significant. 
The transition probability $p_a$ can be approximated by 
\begin{equation}
p_a (r|v_e;\tau)= N(v_e,\tau) r \exp [-(r - m(v_e,\tau))^2/\sigma^2 (v_e,\tau)].
\end{equation}
For small $v_e$ we have $m(v_e,\tau)\sim \alpha(\tau)|v_e|$ and $\sigma^2 (v_e,\tau))\sim \beta(\tau)$. From the functional form of $p_a$ one can see that the transport of the tracer particles is of probabilistic nature. For any fixed $v_e$ the tracers travel different distances during the same time $\tau$. We also see a strong dependence on the condition $v_e$ which can be interpreted as deterministic part of the turbulent transport. This distinguishes it from pure diffusion and directly shows that the widely used Corrsin approximation is violated. In the case of deterministic transport $p_a$ would be proportinal to $\delta(r-m(v_e,\tau))$. A good approximation for $p_b$ is given by
\begin{equation}
p_b(v_p|v_e;\tau) = N(v_e,\tau) \exp [-( u_p -m(v_e,\tau))^2 /\sigma^2(v_e,\tau)] 
\end{equation}
 with $m(v_e,\tau)=\alpha(\tau)\tanh (\beta(\tau) v_e)$ and $ \sigma^2 (v_e,\tau)=\gamma(\tau)(1+\delta(\tau)|v_e|)$. For small $v_e$ we see a strong negative correlation between $v_e$ and $v_p$ (here $\tanh(v_e) \sim v_e$). Both quantities tend to cancel in this case. This negative correlation between the sample-space variables in $p_b$ is connected with the sweeping effect. For a tracer starting with $v_1$  travelling the time $\tau$ without changing its velocity we have $u_{el}=v_1-v_2$ when the velocity at the starting point changes during $\tau$ from $v_1$ to $v_2$. In this case we have $u_p=v_2 - v_1=-u_{el}$. This corresponds to an idealized situation but it gives a hint at the cause of the negative correlations in $p_b$. For larger $v_e$ the correlation decreases. This is captured by the fact that $\tanh(\beta(\tau)v_e) \sim const$ for large $v_e$. For both transition PDFs we observe that for increasing $\tau$ the dependence on their conditions vanishes.
\\
Now we want to turn to the question how the transition PDFs
transform the Eulerian PDF $f_e (v_e;r)$ into the Lagrangian PDF
$f_l (u_l;\tau)$. This process is
depicted in Fig.~\ref{fig:pdfEulerLagrange}. The left part of the
figure shows several  examples of $f_e(v_e;r)$.
Applying $\int dr \ p_a(r|v_e;\tau)$  (see equation
\ref{eq:EulerLagrangeIncrement}) superposes different Eulerian PDFs 
$f_e(v_e;r)$
with different variances leading to the triangular shape in the semi-logarithmic plot
of the new PDF $f_{el}(v_e;\tau)$ (middle of Fig.\ref{fig:pdfEulerLagrange}).
During the transition from $f_{el}(v_e;\tau)$ to $f_l(v_l;\tau)$ the variables $v_p$
and $v_e$ are added to form $v_l$. The previously described
observation that $v_p$ and $v_e$ tend to cancel each other for
small $v_e$ leads  to a stronger weighting of very small $v_l$ so
that the new PDF $f_l(v_l;\tau)$ is strongly peaked  around zero
(right part of Fig.\ref{fig:pdfEulerLagrange}). In contrast to the
center of the distribution the tails seem not to be influenced
significantly by $p_b(v_p|v_e;\tau)$. This is in agreement with
the fact that for large $v_e$ the correlation between $v_e$
and $v_l$ decreases.
\\[1ex]
\textit{Three-dimensional turbulence} To get an impression of the
transition probabilities in three dimensional turbulence we used the
data provided by \cite{biferale05pof,biferale04prl} to calculate the
PDF $p_a(r;\tau)=\int d v_e \ p_a(r|v_e;\tau)$  for different $\tau$.
The result is depicted in Fig. \ref{fig:diff3D}. We see that as in
the two-dimensional case the Lagrangian time scale $\tau$ is related to the Eulerian length scales by a PDF. This well known result shows that in principle it is not possible to connect them by a Kolmogorov type relation like $\tau \sim r/\delta u_r$ \cite{biferale04prl}. In this
relation $\tau$ is a typical eddy turnover time connected to eddies of
length scale $r$.
In our example we have chosen three different values of the time delay
$\tau$ taken from the inertial range. Even when the maximum 
of $p_a$ is at a distance $r$
which lies in the Eulerian inertial range there are significant
contributions from very small and very large $r$. From this observation
we can conclude that due to the turbulent transport the Lagrangian statistics
is influenced by contributions from the Eulerian integral and
dissipative length scales.
\\[1ex]
\textit{Relation to the Multifractal approach}
The characterization of the relation between Eulerian and Lagrangian
PDFs with the help of the conditional probabilities $p_a(r|v_e;\tau)$ (Eulerian to semi-Lagrangian transition) and $p_b(v_p|v_e;\tau)$ (semi-Lagrangian to Lagrangian transition) allows
one to recover a well known multifractal approach for relating
Eulerian and Lagrangian structure function exponents
\cite{biferale04prl}. As a side product, a simpler formula
for the Lagrangian structure function exponents of this multifractal
approach will be obtained. Regarding the translation rule in \cite{biferale04prl} we would have a fixed 
relationship $|\tilde{\bs{x}}| \sim v_e \tau$ ($\delta u_r$ corresponds to $u_e$ in our notation) between the time lag $\tau$ and the distance traveled by a tracer particle during this time. This would correspond to 
choosing $p_a \sim \delta(r - v_e \tau)$ in
(\ref{eq:EulerLagrangeIncrement}). The additional assumption \cite{biferale04prl} $\delta
v_{\tau}\sim \delta u_r$ ($u_l \sim u_e$ in our notation) that the
velocity fluctuations on the time scale $\tau$ are proportional to
the fluctuations on the length scale $r$ could be incorporated in
our framework by choosing $p_b \sim \delta(v_l - v_e)$ 
leading to
\begin{eqnarray}
\label{eq:etolbif}
 f_l(v_l;\tau) &=& \int dv_e \, \delta(v_l - v_e) \int_0^\infty dr \, \delta(r-v_e \tau) f_e(v_e  ; r) \nonumber \\
& =& f_e(v_l;v_l \tau).
\end{eqnarray}
The exponents for the Lagrangian structure function exponents can now
be obtained by making use of the Mellin transform
\begin{equation}
  \label{eq:mellin}
  f_e(v_e, r) = \frac{1}{v_e} \int_{-i \infty}^{i\infty} dn \,
  S_e(n) v_e^{-n}
\end{equation}
with $S_e(n) = A_e(n) r^{\zeta_e(n)}$. Here, we use the same notation
as in \cite{yakhot2006}. Please note, that it will be not neccessary to
know the amplitudes $A_e(n)$ but only the Eulerian scaling exponents $\zeta_e(n)$.
Using Eqn. (\ref{eq:etolbif}) and the Mellin transform we obtain
\begin{equation}
  f_l(v_l; \tau) = \frac{1}{v_l} \int_{-i \infty}^{i\infty} dn \,
   A_e(n) \tau^{\zeta_e(n)} v_l^{\zeta_e(n)-n}.
\end{equation}
This Lagrangian PDF is now inserted into the inverse Mellin transform
to obtain the Lagrangian structure functions
\begin{eqnarray}
  \label{eq:invmellin}
  S_l(n) &=& \int_0^\infty d v_l \, v_l^n f_l(v_l; \tau)  \\
  &=& \int_0^\infty d v_l \,
   \frac{1}{v_l} v_l^n \int_{-i \infty}^{i\infty} dj \,
   A_e(j) \tau^{\zeta_e(j)} v_l^{\zeta_e(j)-j}  . \nonumber
\end{eqnarray}
Now we substitute $j'(j) = j - \zeta_e(j)$, $dj' =
(1-\partial_j\zeta_e(j)) dj$ and denote the inverse function by
$j = j(j')$. Thus we have
\begin{eqnarray}
   S_l(n) &=& \int_0^\infty d v_l \,
  \frac{1}{\delta v_l} v_l^n \int_{-i \infty}^{i\infty} dj' \,
   S_l(j') (\delta v_l)^{-j'}
\end{eqnarray}
with $S_l(j') = \frac{A_e(j)}{1-\partial_j \zeta_e(j)} \tau^{\zeta_e(j)}$ and 
$j' = j - \zeta_e(j)$ and we obtain for the exponents
\begin{equation}
  \label{eq:newbif}
  \zeta_l(n-\zeta_E(n)) = \zeta_e(n)
\end{equation}
It remains to show that this relation (\ref{eq:newbif}) is identical
to the formulas derived in Biferale \textit{et al.}
\cite{biferale04prl}. To see this, we shortly repeat the
multrifractal approach which starts with the Eulerian structure
function exponents
\begin{equation}
  \zeta_e(p) = \inf_h\left( ph + 3 -D_e(h)\right)
  = p h_e^* + 3 - D_e(h_e^*) 
\end{equation}
and $p=D_e'(h_e^*)$. Here $D_e(h)$ is the Eulerian singularity spectrum and $h_e^*$ is the
value where the infimum is achieved. The assumption $r = v_e \tau$ appears now in the denominator of the expression of the Lagrangian structure function exponents
\begin{equation}
  \label{eq:lagbif}
  \zeta_l(p) = \inf_h \left( \frac{ph + 3 -D_e(h)}{1-h}\right).
\end{equation}
From this it follows
\begin{align}
  \zeta_l & (p- \zeta_e(p)) \nonumber \\
		& = \inf_h \left(\frac{(p - p h_e^* - 3 + D_e(h_e^*)) h + 3 - D_e(h)}{1-h}\right).
\end{align}
In order to find the infimum we differentiate with respect to $h$
\begin{equation}
D_e'(h_e^*) - D_e'(h) + D_e(h_e^*) - D_e(h) - D_e'(h_e^*) h_e^* + D_e'(h) h \overset{!}{=} 0
\end{equation}
From this it follows that $h^*_L = h_E^*$ and
\begin{equation}
  \zeta_L(p-\zeta_e(p)) = p h_e^* + 3 - D_e(h_e^*) = \zeta_e(p)
\end{equation}
which recovers (\ref{eq:newbif}). A consequence of (\ref{eq:newbif}) and 
(\ref{eq:etolbif}) is that for a self-similar Eulerian velocity field (as we can find it in 
the two dimensional inverse energy cascade) we should find self similar Lagrangian PDFs. 
As mentioned above this is in contradiction to recent experiments
\cite{rivera2007} and our own numerical simulations \cite{kamps2007}. 
\\[1ex]
\textit{Conclusion and Outlook} We presented a straight forward
derivation of an exact relationship between Eulerian 
 and Lagrangian velocity increment PDFs. For the example of two-dimensional 
forced turbulence we were able to explain how it is possible to observe 
strongly non-Gaussian intermittent 
distributions for the Lagrangian velocity increments. The
two mechanisms in this context are the turbulent transport of the
tracers leading to the mixing of statistics from different length
scales and the velocity change at the starting point of the tracers leading to a further 
deformation of the increment PDF. In comparison we analyzed data from simulations of three-
dimensional turbulence. Similar to the two-dimensional case we demonstrated that Lagrangian time- and Eulerian length-scales are connected
via a transition PDF that varies with the time scale. We where also able to show that the well known multifractal model for the Lagrangian structure functions ia a limiting case of the presented translation rule. 
The next step is to estimate the transition probabilities for three-dimensional
turbulence as well as magnetohydrodynamic turbulence
in order to get a deeper understanding of
the influence of the underlying physical mechanisms, especially the presence
of coherent structures on the
translation process. In this context the question why intermittency
in the Lagrangian picture is stronger in magnetohydrodynamics than
in fluid turbulence \cite{homann06unp}, although the situation is
reversed in the Eulerian picture, will be addressed.
\\[1ex]
\textit{Acknowledgements} We are grateful to H.
Homann, M. Wilczek and D. Kleinhans for fruitful
discussions and acknowledge support from the Deutsche
Forschungsgesellschaft (FR 1003/8-1, FG 1048). We also thank the
supercomputing center Cineca (Bologna, Italy) for providing and
hosting of the data for 3D Turbulence.



\bibliographystyle{alphadin} 

\bibliographystyle{plain}           



\end{document}